\documentclass[aps,prb,reprint,twocolumn,superscriptaddress,showpacs]{revtex4-1}

\usepackage{graphicx}
\usepackage{amssymb}
\usepackage[colorlinks=true,linkcolor=blue,citecolor=blue,urlcolor=blue,filecolor=blue]{hyperref}

\begin{document}

\title{Phonon confinement and plasmon-phonon interaction\\ in nanowire based quantum wells}

\author{Bernt Ketterer} 
\affiliation{Laboratoire des Mat\'{e}riaux Semiconducteurs, Institut des Mat\'{e}riaux, Ecole Polytechnique F\'{e}d\'{e}rale de Lausanne, CH-1015 Lausanne, Switzerland}

\author{Jordi Arbiol} 
\affiliation{Instituci\'{o} Catalana de Recerca i Estudis Avan\c{c}ats (ICREA) and Institut de Ci\`{e}ncia de Materials de Barcelona, CSIC, E-08193 Bellaterra, CAT, Spain}

\author{Anna Fontcuberta i Morral}
\email{anna.fontcuberta-morral@epfl.ch}

\affiliation{Laboratoire des Mat\'{e}riaux Semiconducteurs, Institut des Mat\'{e}riaux, Ecole Polytechnique F\'{e}d\'{e}rale de Lausanne, CH-1015 Lausanne, Switzerland}
\affiliation{Walter Schottky Institut and Physik Department, Technische Universit\"at M\"unchen, Am Coulombwall~3, D-85748 Garching, Germany} 

\date{\today}

\begin{abstract} 
Resonant Raman spectroscopy is realized on closely spaced nanowire based quantum wells. Phonon quantization consistent with 2.4 nm thick quantum wells is observed, in agreement with cross-section transmission electron microscopy measurements and photoluminescence experiments. The creation of a high density plasma within the quantized structures is demonstrated by the observation of coupled plasmon-phonon modes. The density of the plasma and thereby the plasmon-phonon interaction is controlled with the excitation power. This work represents a base for further studies on confined high density charge systems in nanowires.
\end{abstract}

\maketitle

\section{Introduction}

Semiconductor nanowires provide a unique platform to create novel types of nanoscale heterostructures exhibiting quantum phenomena. They offer opportunities for new generations of nanoscale photonic and electronic devices such as multi-quantum well (MQW) well lasers and LEDs\cite{ref1,ref2} high performance transistors\cite{ref3}, or optical modulators.\cite{ref4,ref5} Remarkable features of these non-planar heterostructures compared to conventional planar devices include (i) an improved coupling with light, which makes them better absorbers or emitters respectively in solar cells or light emitting diodes\cite{ref6}, (ii) the possibility of realizing both axial and radial heterostructures, (iii) the chance of combining mismatched materials in axial heterostructures, thanks to an effective radial strain release\cite{ref7} and (iv) the possibility of obtaining two dimensional electron gases around the nanowire core.\cite{ref8}

The importance of semiconductors originates from the ability of tailoring the conductivity (carrier type and concentration) through doping, as well as for the possibility of bandgap engineering.\cite{ref9} This has enabled the fabrication of a manifold of devices: from light emitting diodes and solar cells to many types of transistors. In semiconductor materials, free carriers can be generated by intentional doping, thermal or photo-excitation. Bandgap engineering enables the confinement of carriers in nanoscale heterostructures such as quantum wells (QWs), wires or dots.\cite{ref10,ref11} In QWs, carriers can propagate freely in two directions, while confined in the third one. Collective oscillations of charges or plasmons can be easily generated in QWs, as the small size and the confinement renders easier the generation of a high density plasma.\cite{ref12}  For the observation, it is also an important condition that the plasma remains in the confined structure. It is for this reason that plasmons in the core of nanowires have only been observed in indirect bandgap semiconductor nanowires before.\cite{ref13}

The simultaneous assessment of the structural, electronic, and optic properties of an individual nanowire quantum heterostructure is an extremely challenging task.\cite{ref14,ref15,ref16} Optical spectroscopies, such as photoluminescence and Raman spectroscopy, are probably the most versatile tools to probe structural and functional characteristics simultaneously in a non-destructive and non-invasive way: Photoluminescence spectroscopy can give an insight in the optical properties, Raman spectroscopy can provide information on both the structural and electronic properties through inelastic light scattering from vibrational and electronic excitations, respectively.\cite{ref17}

In polar semiconductors such as GaAs, plasmons can interact with longitudinal optical phonons (LO) through their macroscopic electric field, giving rise to coupled phonon-plasmon modes (LO-P).\cite{ref18} This coupling can be observed in the case where the semiconductor is degenerate\cite{ref19} because the plasmon oscillation frequencies become comparable to those of the LO phonons. The frequency of the LO-P modes depends on the total concentration of free carriers and on the dimensionality and therefore it is a measure of the carrier density in the system. The carriers can also interfere with the phonons giving rise to the Fano effect by creating an asymmetric line shape.\cite{ref20} Such an effect has been observed both for a high doping concentration and after high excitation power densities in both bulk and nanoscale materials such as nanowires (especially silicon).\cite{ref21,ref22,ref23} Though, the Fano effect does not easily allow the quantification of the carrier density. Raman spectroscopy is thus an ideal tool because it provides precious information on the electronic system such as carrier density and mobility without the fabrication of electrical contacts.\cite{ref24,ref25} Indeed, this type of measurements can be of great importance in the characterization of nanowires incorporating more complex structure such as a two (one) dimensional electron gas on the facets, achieved by modulation doping.\cite{ref26,ref27,ref28}

\begin{figure} \includegraphics[width=\columnwidth]{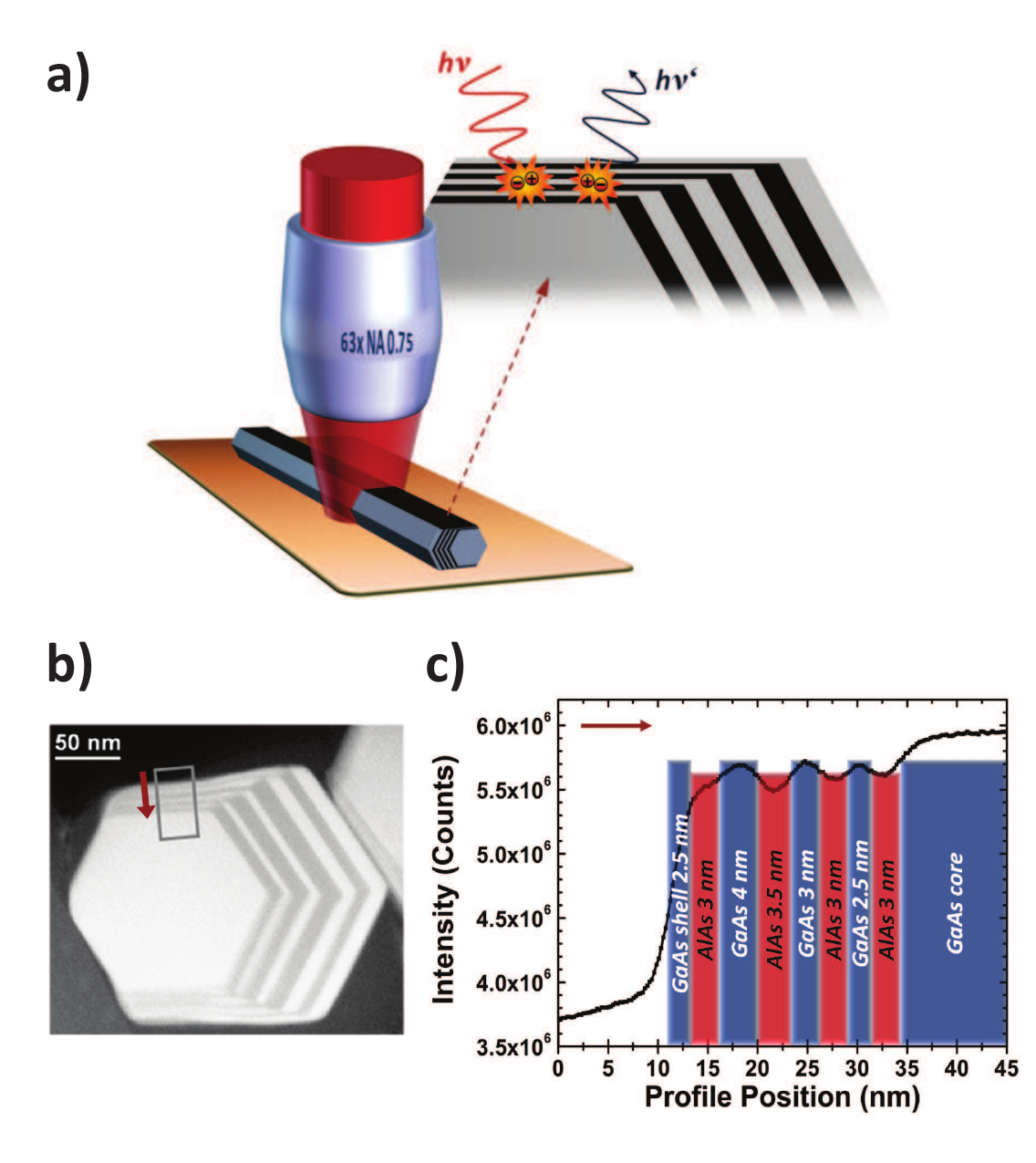}
\caption{\label{figure1}a) Schematic drawing of the experiment where laser light is focused on a single nanowire for resonant Raman spectroscopy in backscattering configuration. Under large excitation power a plasma is formed across the closely spaced MQWs. b) Cross-section TEM micrograph of a GaAs nanowire presenting the AlAs/GaAs MQWs. c) Analysis of the thickness of the MQWs grown on the lateral facets of the nanowire. The intensity contrast is proportional to the Z number and gives information about the composition. We have integrated the intensity in the region marked in the Fig.2b and show the intensity profile following the arrow. For clarity, the arrow is also shown in Fig.2c to indicate the direction in which we show the intensity contrast.}
\end{figure}

In this paper, we report resonant Raman spectroscopy [Fig. \ref{figure1}(a)] performed on an individual nanowire with a shell consisting of a sequence of AlAs/GaAs/AlAs quantum wells [Fig. \ref{figure1}(b)]. The Raman spectra show that the series of closely spaced quantum wells, i.e., the MQW, in the shell of a nanowire is an ideal system to confine a plasma and study the phonon-photon interaction. Here we address prismatic quantum wells where, intentionally, the quantum well thickness changes stepwise between different side facets. Thereby we find the conditions where the high density plasmons can be detected. This is different from typical prismatic or core-shell heterostructures where the quantum well thickness is constant around the nanowire core.\cite{ref29,ref31,ref32}

\section{Experimental}

The GaAs nanowires were grown in a catalyst-free process by using molecular-beam epitaxy (MBE).\cite{ref30} The synthesis was carried out in a high mobility Gen-II MBE system. This allowed us to produce high purity nanowires and grow quantum heterostructures on the nanowire facets with very high crystalline quality and atomically sharp interfaces.\cite{ref31,ref32,ref33} For the growth, two-inch (001) GaAs wafers coated with a sputtered 10 nm thick silicon dioxide were used. The nanowire growth was carried out at a nominal GaAs growth rate of 0.45\AA s$^{-1}$, an As$_4$ partial pressure of 3.5 x 10$^{-6}$ mbar (Ga-rich conditions), a temperature of 630 $^{\circ}$C, and with a rotation of 4 rpm. An analysis of the ensemble revealed that the diameter was 85 $\pm$ 10 nm and uniform along the entire length. High resolution transmission electron microscopy (HRTEM) measurements indicated that the nanowires were single crystalline with a zinc blende structure.\cite{ref34} As the nanowires grow along the [111] direction, the nanowires appeared at an angle of 35$^{\circ}$ with the (001) GaAs substrates. The GaAs nanowires were used as the core for growing a series of multi-quantum wells of AlAs/GaAs with nominal thicknesses of (20/8/25/12/30/18/20 nm) on the facets.  The nominal thickness corresponds to the thickness that would be obtained under such conditions on a planar substrate. Growth of the radial heterostructures on the nanowire facets was achieved by increasing the As$_4$ pressure to 5x10$^{-5}$ mbar. The planar growth rate on the facets is proportional to the magnitude of the flux received. As a consequence, this results in a facet dependent growth rate in the case of nanowires that are non-perpendicular with the substrate.\cite{ref32}  A cross-section TEM image of the nanowire that enables the visualization of the QWs is shown in Fig. \ref{figure1}(b). The thickest quantum wells were obtained for the facets facing the molecular beam flux and the thinnest for the backside facets. For the lateral facets, it has been shown that there is a factor ~3.5-4 between the nominal and measured thickness [Fig. \ref{figure1}(c)].\cite{ref31,ref32} In the following we will label a nanowire with such kind of MQWs on the facets as a MQW nanowire.

Raman scattering experiments were performed in backscattering geometry on single nanowires at 10 K mounted in a cryostat. The 488 nm and 647.1 nm lines of an Ar$^+$Kr$^+$ laser and the 632.8 nm line of a HeNe laser were used for excitation. The laser was focused on the nanowire with a cover-glass corrected objective with numerical aperture NA = 0.75. A xyz-piezostage allowed the positioning of an individual nanowire under the laser spot with a precision of $<$ 2 nm. The power of the incident light was varied between 25 and 500 $\mu$W, corresponding to a change in the laser intensity from 2.9 kW cm$^{-2}$ to 57.5 kW cm$^{-2}$. The scattered light was collected by a TriVista triple spectrometer coupled with a multichannel CCD detector. 

\section{Results and Discussion}
\begin{figure*} \includegraphics[width=0.8\textwidth]{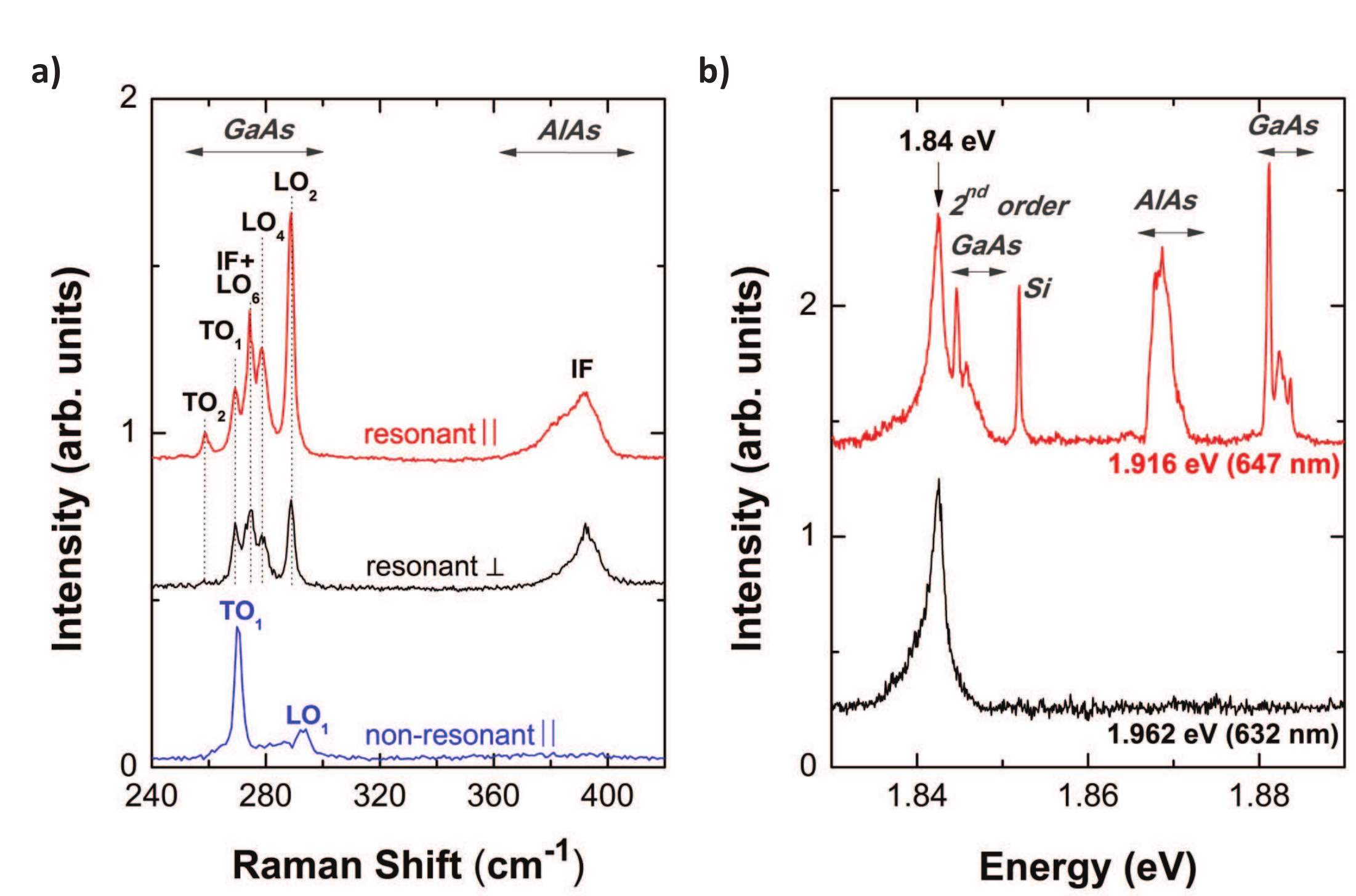}
\caption{\label{figure2}a) Polarized and depolarized spectra taken under resonant excitation with 647.1 nm radiation at very low laser power. For comparison, the spectrum of a the same nanowire under non-resonant excitation with 488 nm radiation is also shown (the intensities have been normalized). b) Spectra of a MQW nanowire taken at 10K with 1.916 eV (647.1 nm) and 1.962 eV (632.8 nm) radiation showing the Raman scattering peaks and the presence of a photoluminescence peak at 1.84 eV.} 		
\end{figure*}

Stokes Raman spectra taken on one of our nanowires are shown in Fig. \ref{figure2}a. We include the measurements taken under parallel and perpendicular polarizations, which correspond to configurations where the polarization of the incident and scattered light are respectively parallel and perpendicular. These configurations are also known as polarized and depolarized. In both cases the incident polarization is parallel to the nanowire axis. For comparison, we have also plotted the polarized spectrum of the same MQW-nanowire taken under non-resonant conditions with 488 nm radiation at the same temperature. The lines between 250 and 300 cm$^{-1}$ correspond to the optical phonon modes of GaAs, i.e. TO, LO, and interface (IF) modes. In the AlAs optical phonon region, the broad peak at 380 cm$^{-1}$ is attributed to in-plane interface modes. In the 250-300 cm$^{-1}$ spectral range, the resonant Raman spectrum of the MQW nanowire is clearly richer than the spectrum taken under non-resonant conditions. The multiple lines appearing close to the TO and LO modes of GaAs can be attributed to quantized optical phonons in the MQWs.\cite{ref35} Indeed, the large difference between the atomic masses of Ga and Al separates the optical phonon bands of the two compounds by more than 100 cm$^{-1}$. Hence, a phonon in one material is heavily damped in the other. As a consequence, the optical phonons with wave vectors normal to the layer planes are confined to the individual MQW layers, which in this way act also as phononic QWs. The quantized wavevectors q allowed in such structures can be calculated in an analogous way to those of a vibrating string with fixed ends\cite{ref36}: 
\begin{equation}
q=\frac{m \pi}{d+\delta}
\end{equation}
where m = 1,2,.. is an integer called the mode order, $d$ is the GaAs quantum well thickness, and $\delta$ is the penetration depth of the phonon mode into the barrier (typically $\delta \sim 1$  monolayer). We have labeled each of the modes appearing in the spectra of Fig. \ref{figure2}a. One should note that only confined longitudinal modes LO$_m$ with even symmetry appear in both the polarized and depolarized spectra. This is a consequence of the symmetry properties of the Fr\"ohlich electron-LO-phonon interaction\cite{ref37}, which is the dominant scattering mechanism when the photon energy is near resonant electronic excitations. In Fig. \ref{figure2}b we show the spectra collected in luminescence mode collected for excitations at 647.1 and 632.8 nm. We observe in both cases a peak at 1.84 eV. Under the excitation of 647.1 nm additional peaks corresponding to Raman scattering appear. Under non-resonant conditions Raman scattering is several orders of magnitude weaker than luminescence. The enhancement of the Raman signal for the excitation at 647.1 nm (1.916 eV) is a clear indication of resonance excitation. Now, to understand the origin of the resonance, one should consider the electronic structure of the MQWs. Due to the small thickness of the GaAs MQWs and the AlAs barriers, minibands form in the conduction and light and heavy hole valence bands. Using the Kronig-Penney model one can calculate the energy position of such minibands. For QW and barrier thickness between 2 and 3 nm, we estimate that the transition energies between the conduction and heavy hole miniband should be about 1.8 eV. This is in good agreement with the observation of luminescence at 1.84 eV. One should add that inherently to the small thickness of the QWs and barriers, the resonance scattering profile should be relatively broad.\cite{ref37} As shown in the work by Cardona et al \cite{ref37}, the resonance profile can be broader than 100 meV. This means that the excitation at 1.916 eV is in strong resonance with the transition between the first conduction and the heavy-hole subband. We should also mention that this luminescence peak exhibits a high degree of polarization, as expected from the antenna geometry of the nanowire. 
It is important to note that, as for the thick MQWs on the other facet of the nanowire and for the nanowire core the resonance conditions do not apply, we exclusively see Raman scattering from the resonantly enhanced thin MQWs. To obtain measurable Raman signal from the thick quantum wells or from the nanowire core, a significant increase of the excitation power or the integration times is necessary. To further proof that the scattering from the narrow MQWs dominates the Raman spectra, we estimate the quantum well thickness from the position of the confined optical modes using equation 1. For this, the phonon wavevectors are related to relevant phononfrequencies via the phonon dispersion of GaAs.\cite{ref38} Comparing these values with the experimentally measured frequencies we find a good agreement for a quantum well thickness of 12 $\pm$ 1 GaAs monolayers, corresponding to 24 $\pm$ 2\AA. The cross-section TEM micrograph of one of the MQW nanowires is shown in Figure \ref{figure1}b. As mentioned above, the thickness of the QWs on the facets should depend on the angle of facets with the molecular beam.\cite{ref32} The thickness of the GaAs QWs on the lateral facets ranges between 2.5 and 3.5 nm, while the AlAs layers have a thickness between 3 and 3.5 nm. The GaAs thickness corresponds quite well with the thickness of the QWs obtained from the LO quantized modes. As a consequence, this measurement allows us to identify which facet of the nanowire is perpendicular to the light beam in the Raman spectroscopy experiment. This is important when we will discuss the high density plasma which we achieved.

\begin{figure} \includegraphics[width=\columnwidth]{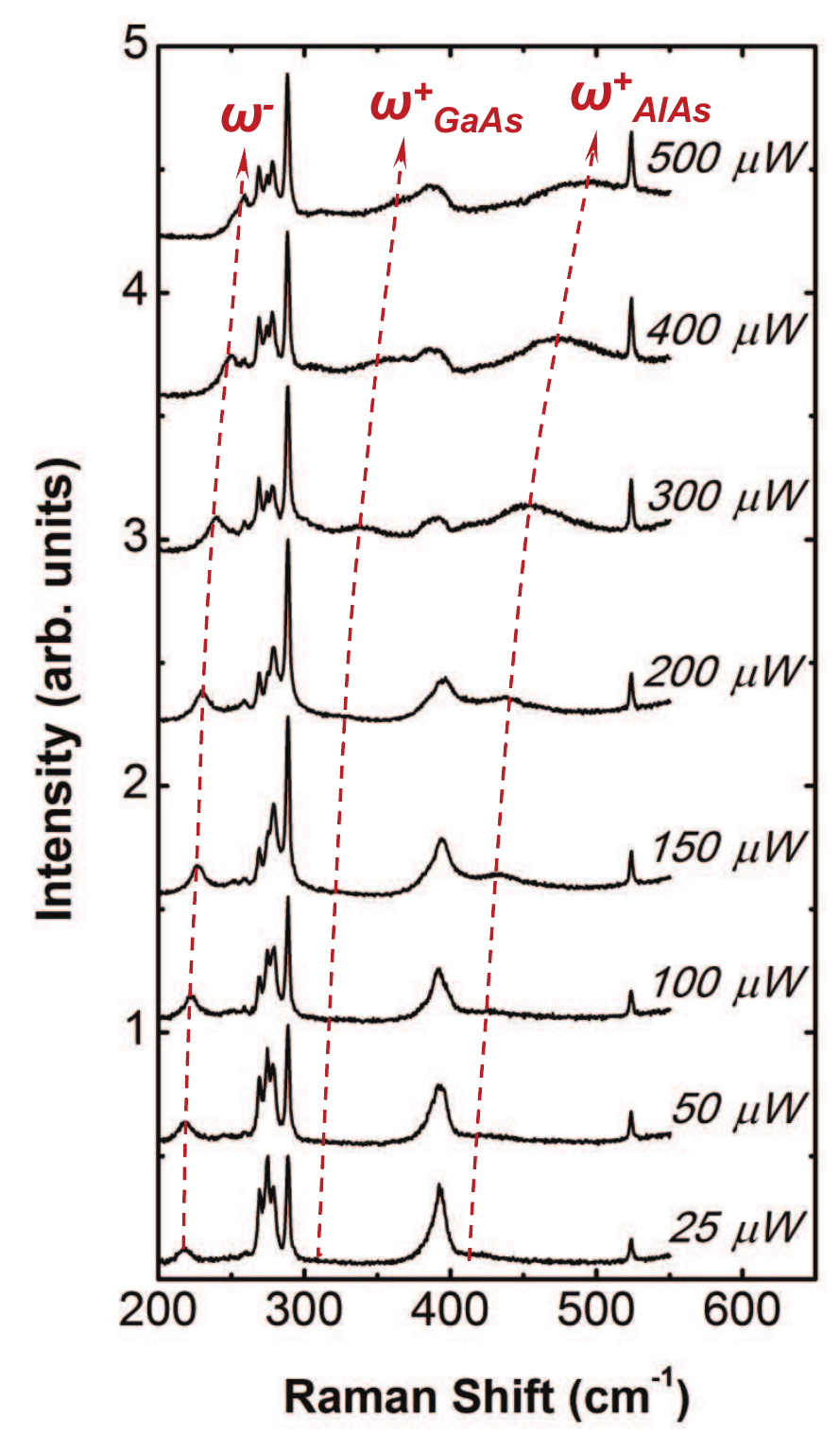}
\caption{\label{figure3}Raman spectroscopy on a single MQW nanowire. The laser intensity is tuned from 2.9 kW cm$^{-2}$ (bottom curve) to 57.5 kW cm$^{-2}$ (top curve). The spectra are offset for clarity.}
 \end{figure}

We now examine the electronic excitations. In isolated quantum wells, carriers can propagate freely in the two in-plane directions and remain confined in the third one, i.e., perpendicular to the plane. The situation can change in closely spaced multi-quantum-well structures. If the barrier layers are thin enough, the overlapping of the electronic wave functions in neighboring wells results in the formation of sub-bands.\cite{ref39} Then, if plasmons are excited in such structures they stop to be purely two-dimensional and propagate both in the in-plane and perpendicular directions of the quantum wells.\cite{ref40} Under sufficient continuous illumination power, a photo-excited steady state plasma can be generated within the quantum wells. Raman spectra taken on a single MQW nanowire for an increasing excitation power are shown in Fig. \ref{figure3}. In addition to the spectral features presented above, we observe the existence of three other peaks that shift to higher energies as the excitation power is increased. We attribute these lines to the phonon-plasmon interaction. Indeed, the interaction of the plasmons with the optical phonon modes of the MQW should result in the creation of three coupled phonon-plasmon modes. One low-frequency $\omega^-$ coupled mode and two high-frequency $\omega^+_{GaAs}$ (GaAs-like) and $\omega^+_{AlAs}$ (AlAs-like ) coupled modes are expected in the Raman spectra.\cite{ref41} The modes shift in frequency for an increasing excitation power as a consequence of the increase in free carrier concentration in the QWs. Here, the free carrier concentration and with it the frequency of the coupled modes can be dynamically tuned in the experiment by changing the laser power.

\begin{figure} \includegraphics[width=\columnwidth]{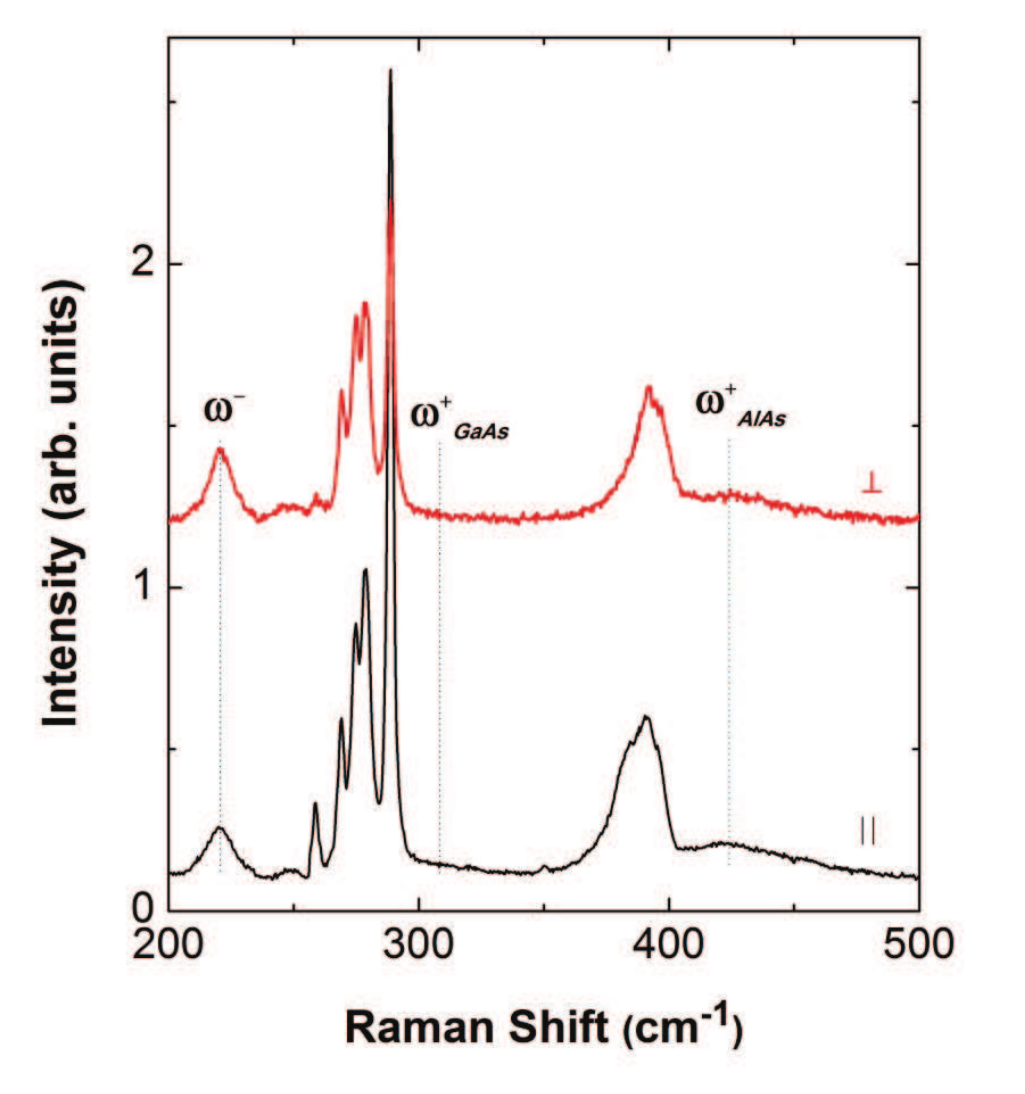}
\caption{\label{figure4}Raman spectra of the coupled phonon-plasmon modes in polarized and depolarized configuration. The excitation power is 100 $\mu$W.}
 \end{figure}

The coupled phonon-plasmon modes are observed in both the polarized and the depolarized Raman spectra (Fig.\ref{figure4}). In the photoexcited MQWs the phonon and electron-hole excitations are renormalized into mixed phonon-plasmon modes. Consequently, the coupled modes can be excited either by the charge-density mechanism or by the forbidden intraband Fr\"{o}hlich interaction. \cite{ref42} Both mechanisms imply polarized scattering with parallel incident and scattered light. However, under strong resonance conditions Fr\"{o}hlich induced LO phonons have been observed in both the polarized and depolarized configuration. To explain this unexpected behavior, an impurity-induced Fr\"{o}hlich interaction was suggested to explain the presence of depolarized scattering. \cite{ref37}. We suggest that this mechanism also contributes to the Raman scattering from the coupled phonon-plasmon modes in the depolarized configuration.  

\begin{figure} \includegraphics[width=\columnwidth]{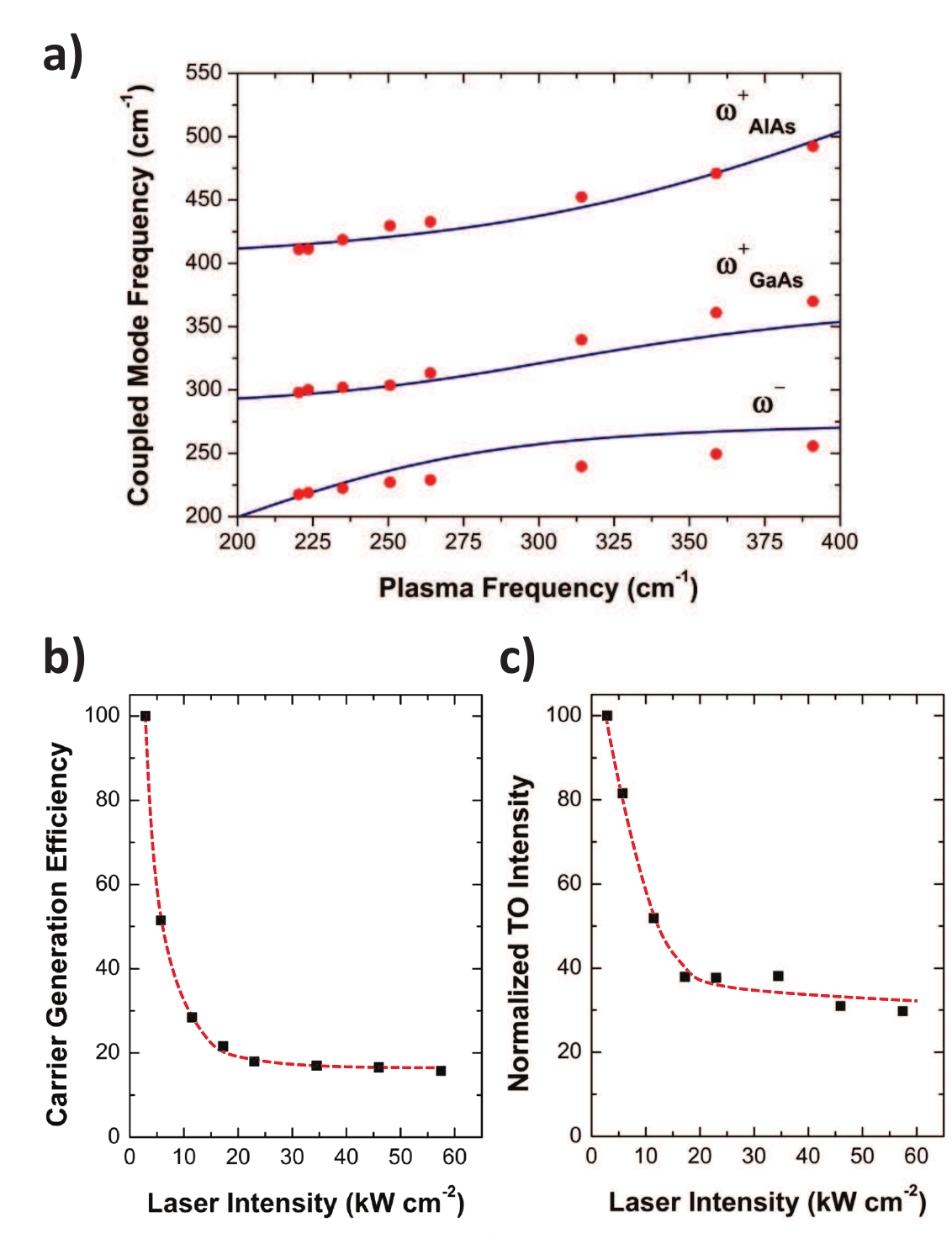}
\caption{\label{figure5}Calculated (lines) and experimental frequencies (points) of the coupled plasmon-phonon modes with the electric polarization directed normal to the layers. b) Carrier generation efficiency (\%) and normalized TO intensity (\%)}
 \end{figure}

We calculated the frequencies of the plasmon-phonon coupled modes propagating along the MQW axis as a function of the excitation power within the framework of the effective medium theory as described elsewhere.\cite{ref41} The results are shown in Fig. \ref{figure5}a. By fitting the experimental data and the theoretical curves we determined the three dimensional plasma frequency $\omega_{Pl}=\sqrt{4 \pi e^2 n/m^*}$  for the intrasubband plasmon vibrations along the MQW axis. We want to note that the contribution of photogenerated holes to the plasma frequency and the overall behavior of the coupled modes can be neglected due to the high effective hole mass. \cite{ref43} The plasma frequency increases from 220.3 cm$^{-1}$ to 391 cm$^{-1}$ when the laser intensity is tuned from 2.9 kW cm$^{-2}$ to 57.5 kW cm$^{-2}$. This corresponds to an increase of the carrier density n from 4.2 to 13.0 x 10$^{17}$ cm$^{-3}$ (with the effective mass $m^*$ = 0.0665 $m_0$ of bulk GaAs). Interestingly, there is only a threefold increase in the carrier density for an increase of excitation power of a factor of nearly 20.

In order to understand this, we calculate the efficiency in the carrier generation as a function of the excitation power. In Fig. \ref{figure5}b we plot the carrier generation efficiency by normalizing it to the efficiency at lowest laser intensity in the experiment. The efficiency decreases with increasing laser intensity and levels off at about 15\%. This effect may be explained by the onset of state filling under strong excitation. Here, further optical transitions into these states become much less probable and the efficiency of carrier generation diminishes. But also the activation of other non-radiative recombination channels such as Auger processes can significantly decrease the carrier generation efficiency at high carrier densities.  
One should point out, that we find the same nonlinear effect in the Raman scattering efficiency. The relative intensity of the TO mode normalized to the corresponding laser intensity is shown in Fig. \ref{figure5}c. It decreases nonlinear with increasing excitation power. The behavior is similar to the decreasing efficiency for photo-excited carrier generation with increasing laser intensity. Accordingly, the probability of a Raman scattering event decreases when either the real intermediate states of the resonant Raman scattering process are less accessible due to the state filling at high excitation, or when other non-radiative recombination paths compete with the Raman scattering event.

\section{Conclusions}
As a conclusion, we have demonstrated the phonon confinement and creation of an electron plasma on closely spaced MQWs grown on the facets of a GaAs nanowire. The excitation power is used to control the plasma density, as shown by the position of the modes of the plasmon-phonon interaction. Efficient doping and the experimental determination of the resulting carrier concentration represent major challenges in current nanowire research. We have shown that resonant Raman scattering can be used as a meaningful probe to measure electron densities in nanowires. Importantly, the plasmon modes discussed here will be powerful tool to evaluate carrier concentrations in doped nanowires incorporating QWs or heterojunctions. 

\begin{acknowledgments}
We thank M Bichler for experimental support in the MBE system. Funding: SNF “Catalyst-free direct doping of MBE grown III-V nanowires”, the ERC starting grant ‘UpCon’, Marie Curie Excellence Grant ‘Senfed’, the DFG through excellence cluster Nanosystems Initiative Munich and SFB 631.
\end{acknowledgments}

\end{document}